\documentclass[conference, 10pt]{IEEEtran}
\IEEEoverridecommandlockouts 
\usepackage[]{graphicx}
\graphicspath{{Figures/}}
\usepackage{caption}
\usepackage{subfig} 
\usepackage{amsmath}
\usepackage{amssymb}
\usepackage{epsfig}
\usepackage{cite}
\usepackage{color}
\usepackage{balance}
\usepackage{amsmath}
\usepackage{amsfonts} 
\usepackage{graphicx}
\usepackage{pdfpages}
\usepackage[T1]{fontenc} 
\usepackage{array}
\usepackage{url}

\usepackage{color}
\usepackage{wrapfig}
\usepackage{lipsum}
\usepackage[hidelinks]{hyperref}
\usepackage{multirow}
\usepackage{tabularx}
\usepackage{tikz}
\usepackage{nth} 
\usepackage{algpseudocode} 
\usepackage{algorithm,algpseudocode}
\usepackage{breqn}
\usepackage{todonotes}
\usepackage{verbatim}
\usepackage{listings}
\usepackage{xcolor}

\definecolor{codegreen}{rgb}{0,0.6,0}
\definecolor{codegray}{rgb}{0.5,0.5,0.5}
\definecolor{codepurple}{rgb}{0.58,0,0.82}
\definecolor{backcolour}{rgb}{0.95,0.95,0.92}

\lstdefinestyle{mystyle}{
    backgroundcolor=\color{backcolour},   
    commentstyle=\color{codegreen},
    keywordstyle=\color{magenta},
    numberstyle=\tiny\color{codegray},
    stringstyle=\color{codepurple},
    basicstyle=\ttfamily\footnotesize,
    breakatwhitespace=false,         
    breaklines=true,                 
    captionpos=b,                    
    keepspaces=true,                 
    numbers=left,                    
    numbersep=5pt,                  
    showspaces=false,                
    showstringspaces=false,
    showtabs=false,                  
    tabsize=2
}
\begin{document}
\title{Automated Climate Analyses Using \\Knowledge Graph}

\author{\IEEEauthorblockN{
Jiantao~Wu\IEEEauthorrefmark{1}\IEEEauthorrefmark{2},
Huan~Chen\IEEEauthorrefmark{2},
Fabrizio~Orlandi\IEEEauthorrefmark{1}\IEEEauthorrefmark{3},
Yee Hui Lee\IEEEauthorrefmark{4},
Declan O'Sullivan\IEEEauthorrefmark{1}\IEEEauthorrefmark{3}, and
Soumyabrata Dev\IEEEauthorrefmark{1}\IEEEauthorrefmark{2}
}
\IEEEauthorblockA{\IEEEauthorrefmark{1} ADAPT SFI Research Centre, Dublin, Ireland}
\IEEEauthorblockA{\IEEEauthorrefmark{2} School of Computer Science, University College Dublin, Ireland}
\IEEEauthorblockA{\IEEEauthorrefmark{3} School of Computer Science and Statistics, Trinity College Dublin, Ireland}
\IEEEauthorblockA{\IEEEauthorrefmark{4} School of Electrical and Electronic Engineering, Nanyang Technological University (NTU), Singapore}

\thanks{This research was partially funded by the EU H2020 research and innovation programme under the Marie Skłodowska-Curie Grant Agreement No.~713567 at the ADAPT SFI Research Centre at Trinity College Dublin. The ADAPT Centre for Digital Content Technology is funded under the SFI Research Centres Programme (Grant 13/RC/2106\_P2) and is co-funded under the European Regional Development Fund.}
\vspace{-0.6cm}
}

\maketitle

\begin{abstract}
The FAIR (Findable, Accessible, Interoperable, Reusable) data principles are fundamental for climate researchers and all stakeholders in the current digital ecosystem. In this paper, we demonstrate how relational climate data can be "FAIR" and modeled using RDF, in line with Semantic Web technologies and our Climate Analysis ontology. Thus, heterogeneous climate data can be stored in graph databases and offered as Linked Data on the Web. As a result, climate researchers will be able to use the standard SPARQL query language to query these sources directly on the Web. In this paper, we demonstrate the usefulness of our SPARQL endpoint for automated climate analytics. We illustrate two sample use cases that establish the advantage of representing climate data as knowledge graphs. 

\end{abstract}

\IEEEpeerreviewmaketitle

\section{Introduction}
Knowledge Graphs (KG) have successfully contributed to a variety of applications, including information retrieval, chatbots\cite{liang2018knowledge}, smart energy systems~\cite{orlandi2019interlinking}. In this work, we explore how to perform weather analysis by querying our climate knowledge graph\footnote{\url{http://jresearch.ucd.ie/linkclimate/}}. The data in our climate knowledge graph is modeled in RDF (Resource Description Framework) following our Climate Analysis (CA) ontology~\cite{wu2021ontology} and is stored in a SPARQL endpoint. The creation of climate KGs provides an interoperable data structure for climate data, which makes the data easily integrated with other domain data modeled by other ontologies across federated endpoints. We use the data from the National Oceanic and Atmospheric Administration (NOAA), a reputable climate data center that provides raw tables (\textit{e.g.} RDB, CSV, JSON) of global climate datasets such as temperature and precipitation~\cite{manandhar2018importance}. The rest of the paper begins by outlining the context for this work, followed by some examples of automated 
climate analysis. Finally, we conclude the paper describing our future work.

\section{Representing Climate Data Using KG}
In this section, we include context information about the initial data source NOAA Climate Data and our Climate Analysis Ontology.
\paragraph{NOAA Climate Data}
The National Oceanic and Atmospheric Administration (NOAA) is a research organization that studies the seas, major rivers, and atmosphere
\cite{center2012national}. NOAA Climate Data Online (CDO) offers free access to the National Climatic Data Center's (NCDC) database of global historical weather and climate data, as well as station history data\cite{graham2014growing}. The NOAA's RESTful APIs provide access to data in CSV, SSV, JSON, PDF, and NetCDF formats. Our work is based on NOAA climate data\cite{difranzo2011web}.

\paragraph{Climate Analysis Ontology}
In our research, we use the CA ontology~\cite{wu2021ontology} to model the NOAA daily summary datasets. There are two major classes in the CA ontology: (\textit{i}) Station and (\textit{ii}) Observation. A station is a site that generates climate observations. Geographical factors (longitude and latitude) or literal names can be used to identify a station. An observation is used to describe a climate feature of interest that has been detected that generated by station, the observed value, and the datatype that defines the feature of interest are typically associated with it.

\section{Analyzing Climate Data using 
KG}
This section showcases how to write SPARQL queries to gather climate data for study\footnote{In the spirit of reproducible research, all the source code is available at \url{https://github.com/futaoo/sparql-weatheranalysis}.}. The first sub-section lists example SPARQL queries that are executed for producing data sources used by the climate analysis experiments in the second sub-section.

\subsection{Data Collection}
Listing 1 is the SPARQL query to gather data on the average temperature in Dublin from 1980 to 2019 and Listing 2 is to gain the weather types data in Sculthorpe, England, from 1951 to 1963\footnote{For this older time period, weather types (\textit{e.g} fog, haze) are available for analysis. These are categorical data for which the detailed processing procedure is given in Section~\ref{sec:analysis}.}.



\vspace{-0.3cm}
\begin{figure}[htb]
\centering
\includegraphics[width=0.47\textwidth]{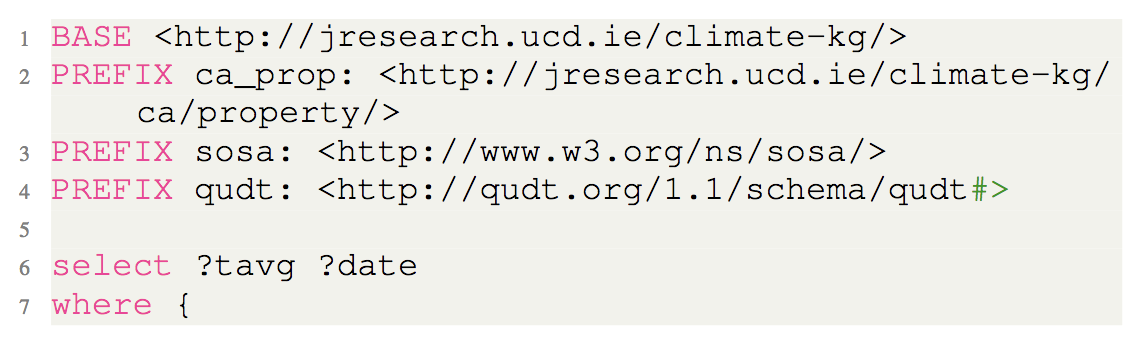}
\end{figure}

\begin{figure}[htb]
\centering
\includegraphics[width=0.49\textwidth]{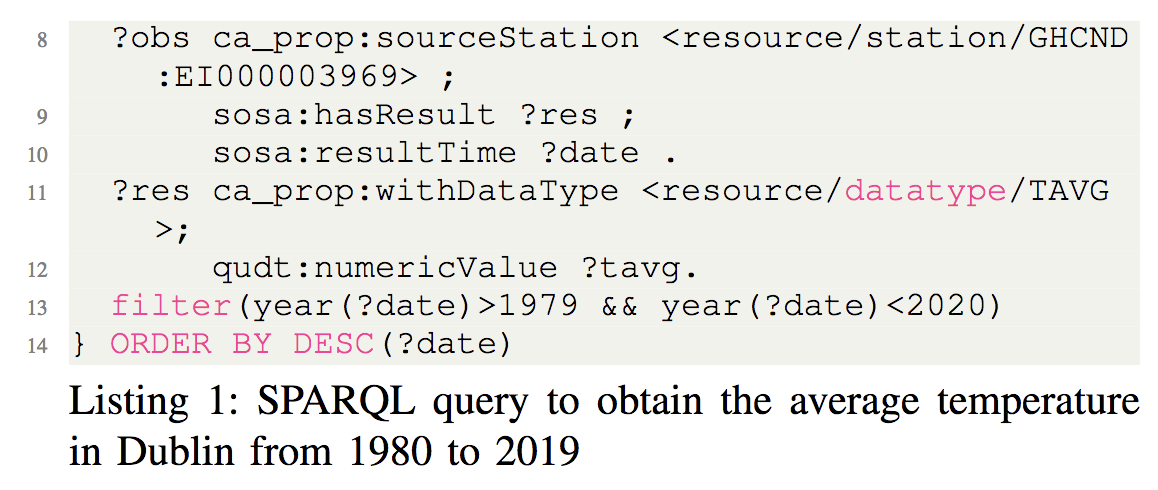}
\end{figure}

\begin{figure}[htb]
\centering
\includegraphics[width=0.49\textwidth]{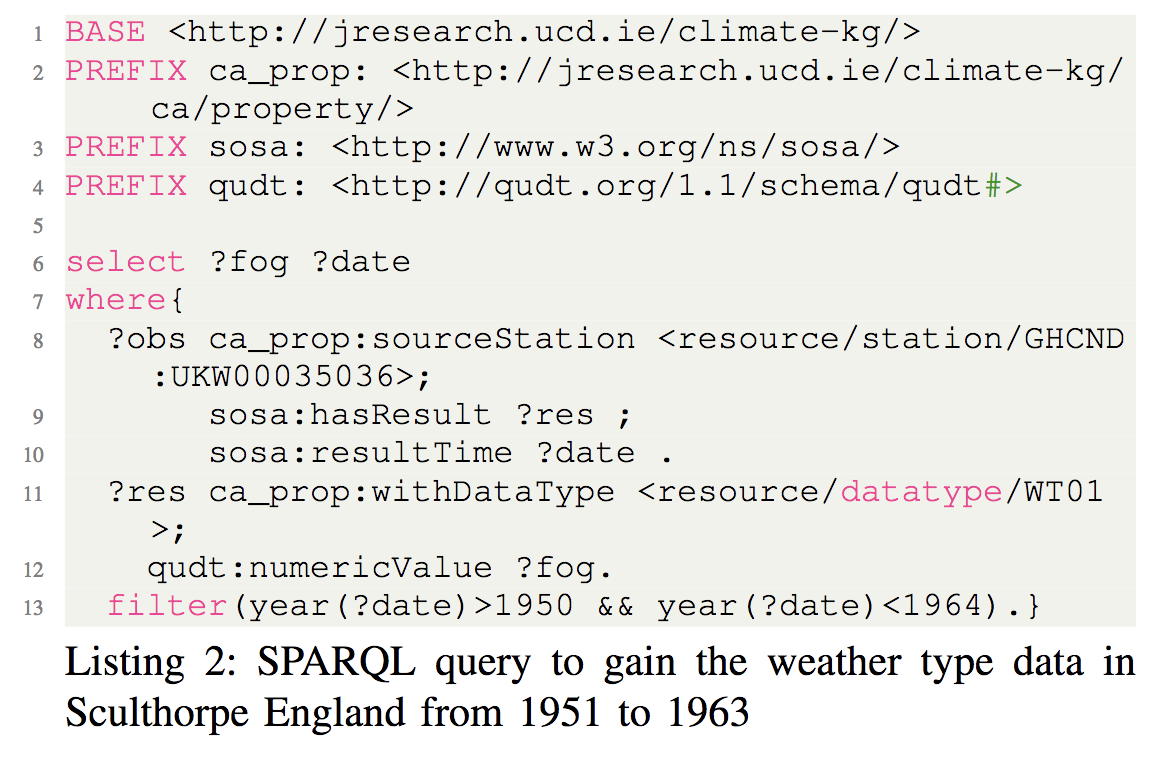}
\end{figure}



\subsection{Use Case I: Variation of monthly temperature}
\label{sec:analysis}
\begin{figure}[ht!]
    \centering
    \includegraphics[width=0.46\textwidth]{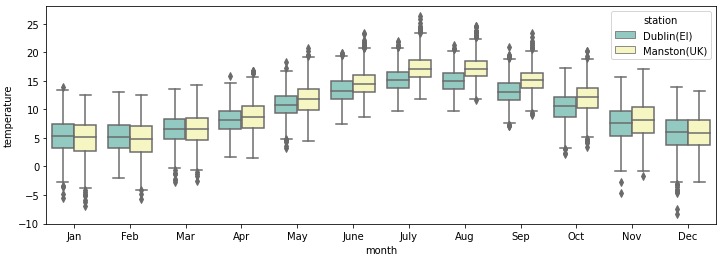}
    \caption{Monthly temperature in Dublin and Manston from 1980 to 2019}
    \label{fig:tprt}
\end{figure}

To analyze the temperature in Dublin (IE) and Manston (UK) from 1980 to 2019, we construct a station-grouped box-plots based on the results (see Fig.~\ref{fig:tprt}). Each packet contains monthly statistics for 40 years daily temperature records. The figure shows that the temperatures in Dublin and Manston reached their highest levels in July and are close in the winter from November to January but in other seasons temperature in Manston is generally a bit higher than Dublin.

To analyze the past weather in Sculthorpe, England, we requested the weather type data 
and used temperature and precipitation data as auxiliary data from our SPARQL endpoint to judge each day's weather condition. The  assumption is that a raining day is determined if the data field showing rainfall is accurate and the day's precipitation is greater than 0 and the temperature is greater than 0. Similar decisions are made on snow that when the field describing snowfall is true and the precipitation is greater than 0 and the temperature is low, then we consider the day to be a snowy day. Drizzle day is asserted when the weather type haze is presented as 1. Additionally, a day without data indicating such weather conditions implies that it was a sunny day.

\subsection{Use Case II: Distribution of weather type}
\begin{figure}[ht!]
\centering
\includegraphics[width=0.48\textwidth]{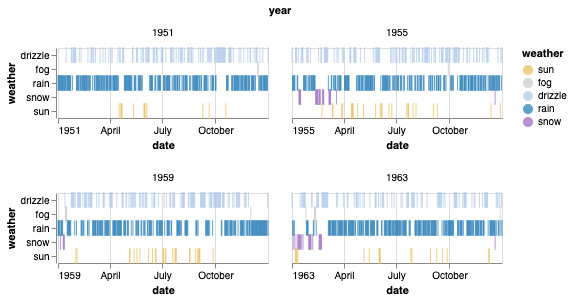}
\caption{Distribution of days by weather type in Sculthorpe, UK}
\label{fig:wtype}
\end{figure}

Finally, we present the strip plot (Fig.~\ref{fig:wtype}) to present the distribution of weather types (rain, fog, drizzle, snow, sun) on every day of the year 1951, 1955, 1959 and 1963. The weather in Sculthorpe was bad most of the time during 1950s as a result of frequent drizzle, fog, and haze, as seen in Fig.~\ref{fig:wtype}.

\section{Conclusion \& Future Work}
In this paper, we demonstrated how to use standards like RDF and SPARQL to obtain knowledge graph data modeled by the CA ontology for climate studies. According to the data type variety, this paper provides two different weather analyses of which the one is based on the statistics on the quantity data about temperature and the other is based on the categorical meteorological records. Both analyses provide basic understanding to the weather conditions in the corresponding cities (\textit{i.e.} Dublin, Manston and Sculthorpe). In the future, we will leverage the power of semantic approaches to enhance the usability of our climate knowledge graph for climate researchers. For instance, integrating additional data sources for more advanced climate studies.





\bibliographystyle{IEEEtran.bst}

\end{document}